# Market Reactions and Information Spillovers in Bank Mergers: A Multi-Method Analysis of the Japanese Banking Sector

Haibo Wang

A.R. Sánchez Jr. School of Business, Texas A&M International University, Laredo, TX 78041, USA; hwang@tamiu.edu

Takeshi Tsuyuguchi

Otterbein University, Westerville, OH, 43230, USA, tsuyuguchi1@otterbein.edu

**Abstract**

Major bank mergers and acquisitions (M&A) transform the financial market structure, but their valuation and spillover effects remain open to question. This study examines the market reaction to two M&A events: the 2005 creation of Mitsubishi UFJ Financial Group following the Financial Big Bang in Japan, and the 2018 merger involving Resona Holdings after the global financial crisis. The multi-method analysis in this research combines several distinct methods to explore these M&A events. An event study using the market model, the capital asset pricing model (CAPM), and the Fama-French three-factor model is implemented to estimate cumulative abnormal returns (CAR) for valuation purposes. Vector autoregression (VAR) models are used to test for Granger causality and map dynamic effects using impulse response functions (IRFs) to investigate spillovers. Propensity score matching (PSM) helps provide a causal estimate of the average treatment effect on the treated (ATT). The analysis detected a significant positive market reaction to the mergers. The findings also suggest the presence of prolonged positive spillovers to other banks, which may indicate a synergistic effect among Japanese banks. Combining these methods provides a unique perspective on M&A events in the Japanese banking sector, offering valuable insights for investors, managers, and regulators concerned with market efficiency and systemic stability.

**1. Introduction**

Bank consolidation is driven by potential value creation through economies of scale and improved efficiency (Beccalli & Frantz, 2013; Houston et al., 2001). At the same time, this trend raises important questions for policymakers about competitive dynamics and potential increases in systemic risk (Beck et al., 2006; Cowan et al., 2022). The Japanese banking sector offers an exciting setting for studying these issues. This study focuses on two mergers and acquisitions (M&A) events: the 2005 M&A of Mitsubishi Tokyo Financial Group and UFJ Holdings, and the 2018 M&A involving Resona Holdings. The M&A decision, driven by regulatory pressures and strategic goals, is a non-random selection (Aranha et al., 2024), which presents a challenge for traditional analytical tools, such as event study or difference-in-differences (DiD), as their assumptions, including parallel trends, may not hold (Baker et al., 2025).

The multi-method approach combines several methods grounded in three research streams to address this challenge. The first method is the extensive work on event studies in finance (Fama et al., 1969). Studies of M&A using this method generally find that shareholders of target firms expect positive cumulative abnormal returns (CAR), while the returns for acquiring firms are more mixed (Andrade et al., 2001). The Granger causality test, based on vector autoregression (VAR) models, can assess information spillovers from an M&A event (M. H. Song & Walkling, 2000). These results can suggest whether returns of M&A banks help predict future returns of their competitors, a technique frequently used to detect spillovers in financial markets (Billio et al., 2012). Finally, the study incorporates techniques from causal inference methods to isolate the causal impact of M&A events using propensity score matching (PSM) (Rosenbaum & Rubin, 1983). PSM is a quasi-experimental technique for constructing a more comparable control group by matching treated and control units on a set of observable characteristics before the event. By comparing outcomes between the treated group and their matched counterparts, it is possible to estimate the average treatment effect on the treated (ATT) to evaluate the M&A's causal impact (Blonigen & Pierce, 2016).

This research answers two main research questions. 1) What was the immediate stock market reaction of these M&A initiation days - the first days of listing on the Tokyo Stock Exchange - on the banks involved? 2) Did the events create spillover effects for competitor banks, and what can be said about their timing and scale? To explore these questions, the multi-method analysis first captures the market's initial reaction, then examines dynamic spillovers to competitors, and finally provides a causally informed estimate of the sustained impact.

## 2. Data and Methodology

### 2.1 M&A Events and Data Sources

The analysis focuses on two M&A events: the merger forming Mitsubishi UFJ Financial Group (MUFG) on October 3, 2005 (U.S. Securities and Exchange Commission, 2005), and the acquisition of Resona Holdings on April 1, 2018 (Resona Holdings, 2018). From the late 1990s to the early 2000s, Japan experienced a sharp rise in non-performing loans, largely due to the bursting of the asset bubble. Japan began deregulating and liberalizing its financial sector in the late 1990s and into the 2000s, a period known as the "Financial Big Bang", causing many banks to be consolidated, merged, or restructured to improve health and reduce the number of failing or weak banking entities (Hoshi & Kashyap, 2004). The MUFG M&A created the world's largest bank by assets at the time, driven by a desire for global competitiveness (Uchino & Uesugi, 2022). Its scale was globally significant, representing a major shock to domestic and international banking systems. After the global financial crisis, large-scale consolidation among Japan's major city banks largely subsided. Since then, most merger activity has been concentrated among smaller regional and local banks, which continue to consolidate in response to demographic pressures and low profitability (Hoshi & Kashyap, 2004). The Resona M&A was highly significant for domestic and regional markets, reflecting survival and an efficient strategy in a low-growth, low-margin operating environment (Onji et al., 2012). The MUFG M&A occurred in a period of recovery from Japan's "lost decade" but before the global financial crisis of 2008. The Resona M&A occurred during the era of "Abenomics" and the Bank of Japan's (BoJ) unprecedentedly low-interest-rate policy (Katsu, 2020). This contrast helps explore whether market reactions and spillover dynamics are sensitive to macroeconomic conditions and monetary policy. The different environments of the two events may have led to various market reactions due to the investors' expectations. Therefore, both M&A events are significant for the evolution of the Japanese interbank market structure (Wang, 2024).

Daily stock price data for the treatment banks (those involved in the M&A) and peer control banks are collected from financial data providers, including Bloomberg and Google Finance. The details of the M&A events are sourced from the BoJ, the Japanese Bankers Association (JBA), and the Financial Services Agency (FSA) (Japanese Bankers Association, 2025). For each event, a 60-day estimation window is tested before a 30-day event window for the time-series and matching analyses.

## 2.2 Empirical Framework

This study proposes a multi-method framework to analyze the impact of the M&A events, as shown in Figure 1.

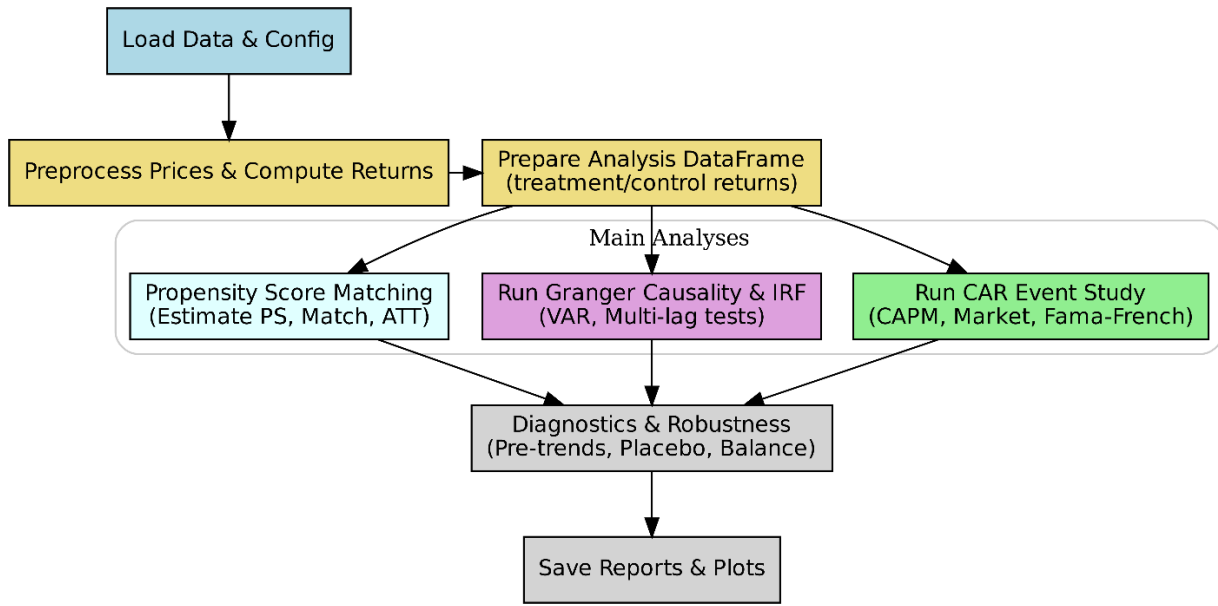

**Figure 1:** Flowchart of the Experimental Design (Authors' Compilation)

Source: Authors' compilation.

Three event study models (the market model (MM), capital asset pricing model (CAPM), and the Fama-French three-factor model) are selected to measure the immediate wealth effect. A Granger causality test is conducted to examine information spillovers between the treatment and control portfolios. Impulse response functions (IRFs) from the VAR are also analyzed to see how a shock to one portfolio's returns affects the other over time. A key difficulty in evaluating M&A is that the choice to merge is not a random event, which can bias the results of simpler comparative methods. To address this selection bias, the study employs PSM to create a control group of non-merging banks that, based on observable characteristics, closely resembles the group of merging banks prior to the event. This allows for the estimation of the ATT. The ATT is the average difference in post-event returns between these matched pairs. This matching approach is increasingly used in corporate finance to strengthen causal claims in M&A research (Masulis et al., 2007; W. Song et al., 2013). The mathematical formulations of the MM, CAPM, Fama-French three-factor, Granger causality test, and PSM methods are provided in the Appendix.

## 3. Empirical Results and Discussions

### 3.1 CAR Event Study Results

Figure 2 and Table 1 illustrate a positive market reaction to the 2018 Resona Holdings M&A, with the CAR for the merging banks exhibiting an upward trend following the event date. Before the event day, the CAR hovers near zero, suggesting that investors were uncertain about the consequences of the merger even though information about the merger had been discussed and made publicly available before the merger, as documented in previous studies (Fama, 1970; Schwert, 1996). After the event, the CAR sustained an upward trend throughout the post-event window, suggesting a lasting positive revaluation of the merging bank rather than just a brief speculative spike. All three models confirmed the findings, indicating the observed outperformance was a genuine reaction to the merger news and not an artifact of broader market trends.

The market's reception of the 2005 MUFG merger in Table 2 was exceptionally positive, indicating a major value-creating event for the shareholders involved. An interesting feature of this event, as seen in Figure 3, is the significant run-up in the CAR before the official event date. This pattern could indicate information asymmetry regarding the worth of the acquisition premium, as suggested in some of the prior literature (Andrade et al., 2001), as well as the market anticipated information about the impending deal. The sheer scale of the abnormal returns signals that investors saw the creation of a global-scale financial institution as a positive strategic move.

**Table 1:** Event Study Results for the Resona Holdings M&A

| Model | Mean Daily AR (%) | T-statistic | CAR (%) |
|---|---|---|---|

| Model | Mean Daily AR (%) | T-statistic | CAR (%) |
|---|---|---|---|
| CAPM | 0.18 | 5.02*** | 5.64 |
| Market Model | 0.23 | 6.22*** | 6.99 |
| Fama-French Three-Factor | 0.18 | 5.04*** | 5.66 |

Note: *** p < 0.01, ** p < 0.05, and * p < 0.10*.

**Table 2:** Event Study Results for the MUFG M&A

| Model | Mean Daily AR (%) | T-statistic | CAR (%) |
|---|---|---|---|
| CAPM | 0.69 | 7.05*** | 29.79 |
| Market Model | 0.27 | 2.84*** | 12.01 |
| Fama-French Three-Factor | 0.64 | 6.48*** | 27.37 |

Note: *** p < 0.01, ** p < 0.05, and * p < 0.10*.

**Figure 2:** CAR results of the Resona Holdings M&A event on April 1, 2018

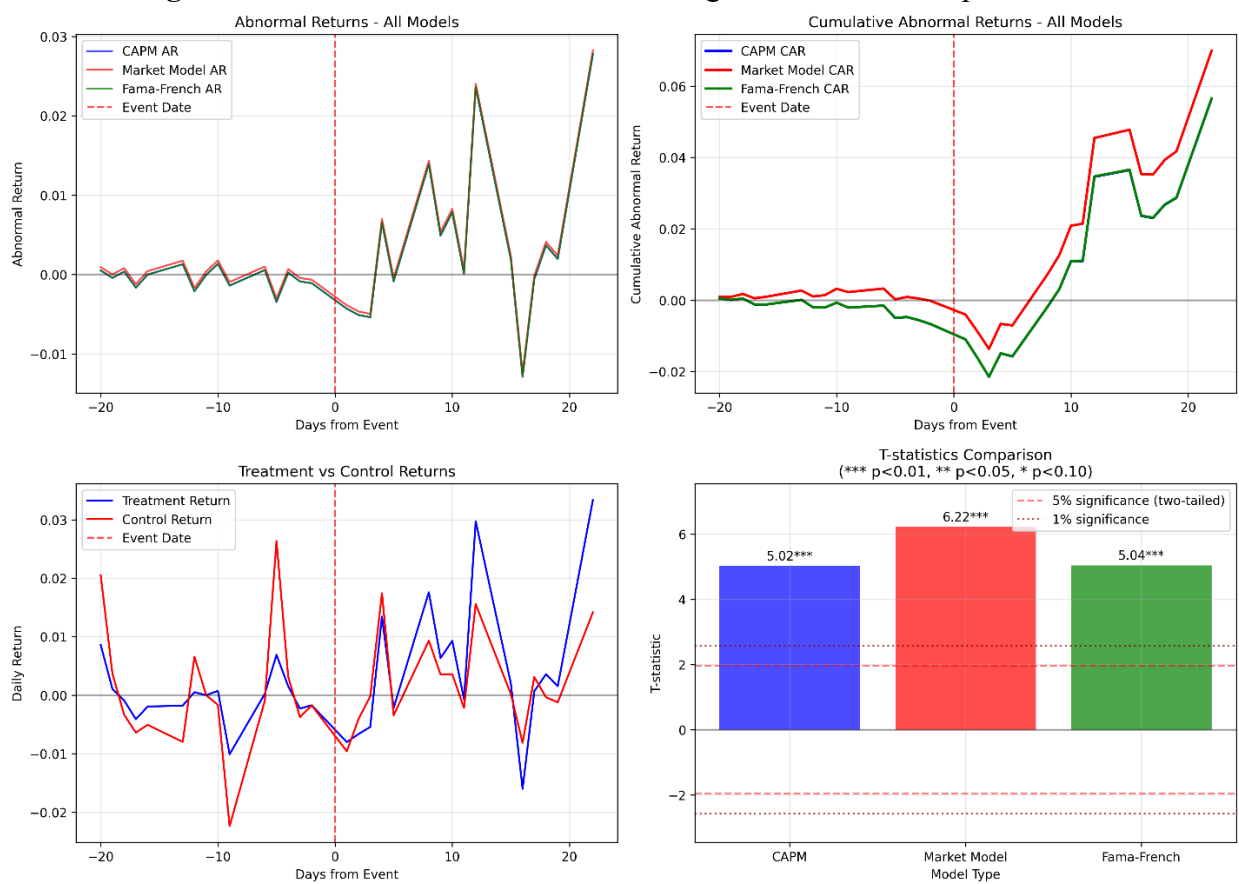

Note: The figure presents the event study analysis for the Resona Holdings merger announced on April 1, 2018. The top-right panel shows the CAR for the treatment portfolio, calculated using three different models. All models indicate a positive market reaction. The final CAR using the market model was +6.99%, while the more conservative Fama-French model yielded a CAR of +5.66%. The bottom-right panel confirms that all models' mean abnormal returns were statistically significant, with t-statistics ranging from 5.02 to 6.22 (all $p < 0.01$). The top-left panel shows the daily abnormal returns, while the bottom-left panel compares the raw daily returns of the treatment portfolio to the control portfolio.

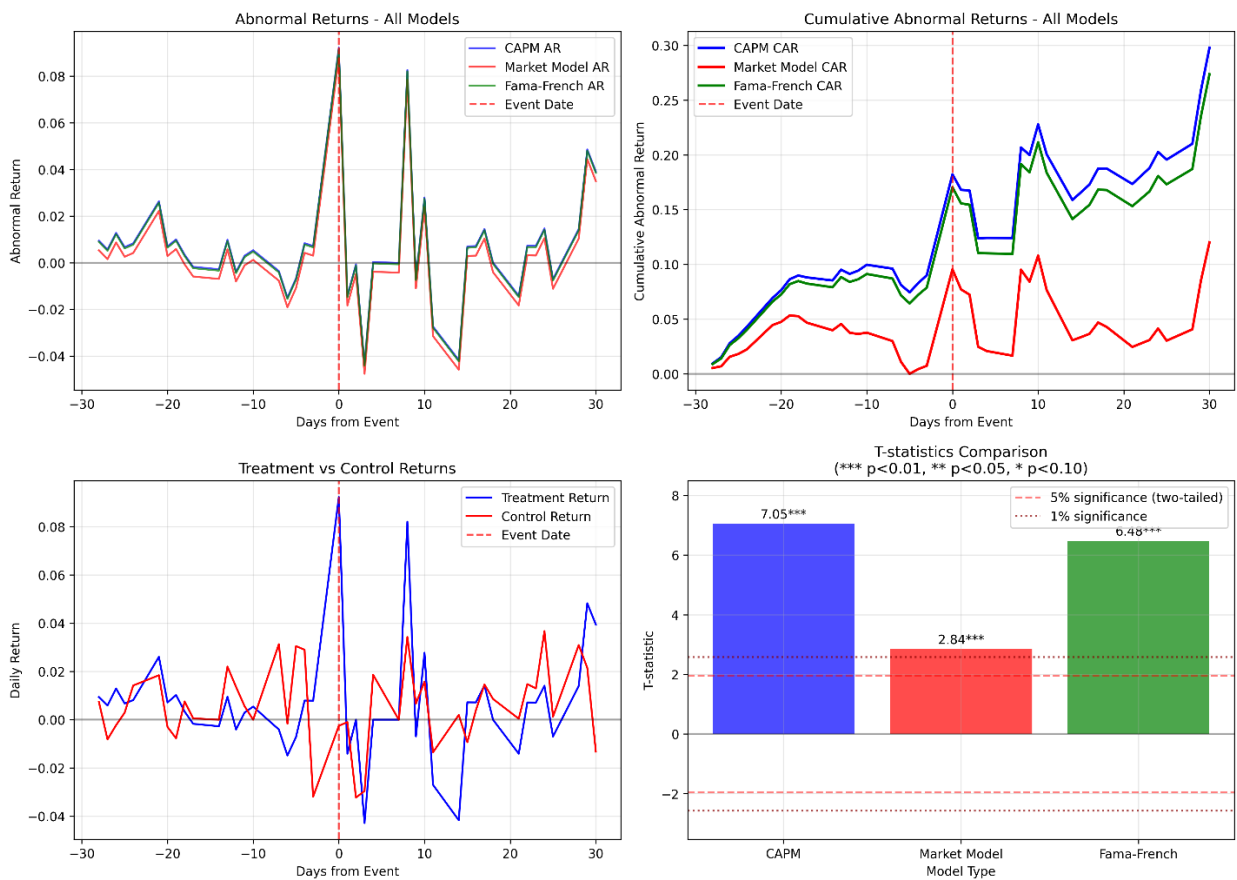

**Figure 3:** CAR results of the MUFJ M&A event on October 3, 2005

Note: The figure displays the event study results for the MUFG merger event on October 3, 2005. The top-right panel shows the CAR for the treatment portfolio. The reaction was overwhelmingly positive and statistically significant across all models. The Fama-French three-factor model estimates a final CAR of +29.25%, while the CAPM estimates a CAR of +29.72%. While still showing a significant effect, the standard market model produced a more conservative CAR of +13.55%. The bottom-right panel shows that the mean abnormal returns were highly significant, with t-statistics for the F-F and CAPM models exceeding 7.5 ($p < 0.01$).

## 3.2 Granger Causality and Spillover Results

Based on the market reaction to the M&A, the VAR model can detect the dynamic interplay between the merging banks and their competitors. The results from the Granger causality tests point toward an apparent, one-way information spillover from the treatment portfolio to the control portfolio. The tests in Figure 4 show that the past returns of the merging banks have statistically significant predictive power over the future returns of their competitors, particularly at a two-day lag (F-statistic = 3.80, p = 0.025 in Table 3). This pattern seems to suggest that the market first responds to the news directly affecting the merging firms, and this information then spreads to the rest of the industry as investors reassess the competitive landscape.

Granger causality tests reveal a persistent, one-way spillover from the treatment portfolio to the control portfolio. This predictive relationship is statistically significant at multiple time horizons, with the strongest effect appearing at a lag of three days (F-statistic = 3.26, p = 0.023 in Table 4) but also showing significance at a lag of one day. This pattern in Figure 5 may suggest a multi-day market adjustment process, where investors first react to the initial news and then update their valuations of competitor firms as the full strategic implications of creating a global megabank become clearer.

**Figure 4:** IRF results of the Resona Holdings merger on April 1, 2018

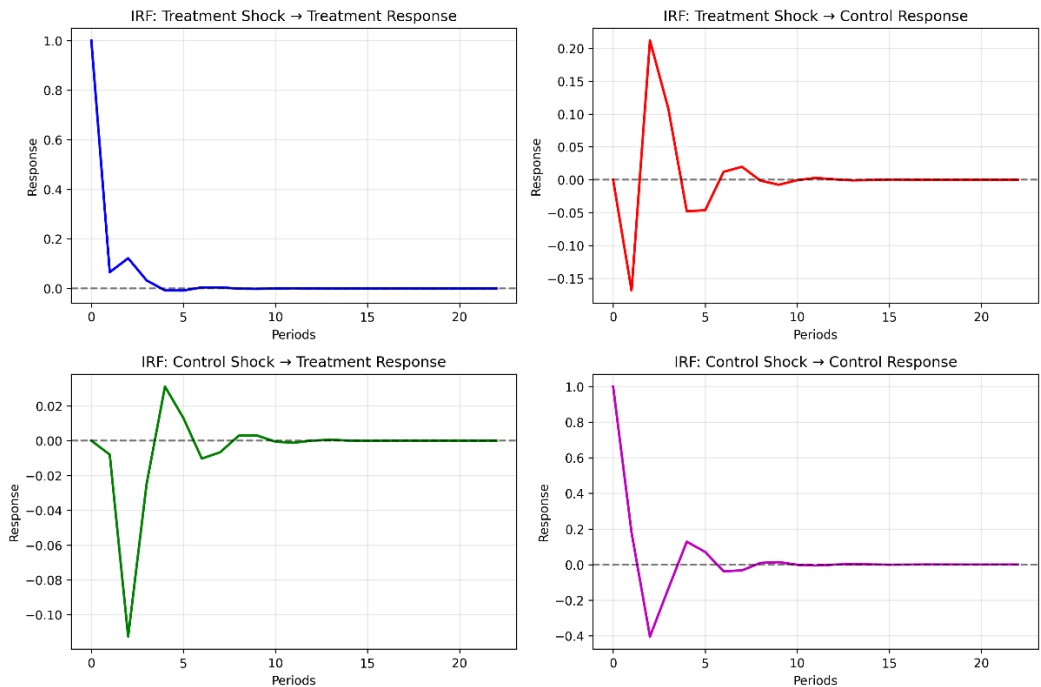

Note: The figure displays the IRFs from a VAR model with a lag of 2, showing the dynamic response of each portfolio to a one-standard-deviation shock in the other. The most critical panel is the top right, which shows the response of the control portfolio to a shock in the treatment portfolio. The plot indicates that a positive shock to the merging banks' returns is followed by a significant positive response to the competitor banks' returns one period later, an effect that quickly dissipates. The top-left and bottom-right panels show each portfolio's expected response to its shock, which dies out rapidly. The bottom-left panel shows the insignificant response of the treatment portfolio to a shock in the control portfolio, consistent with the Granger causality test results.

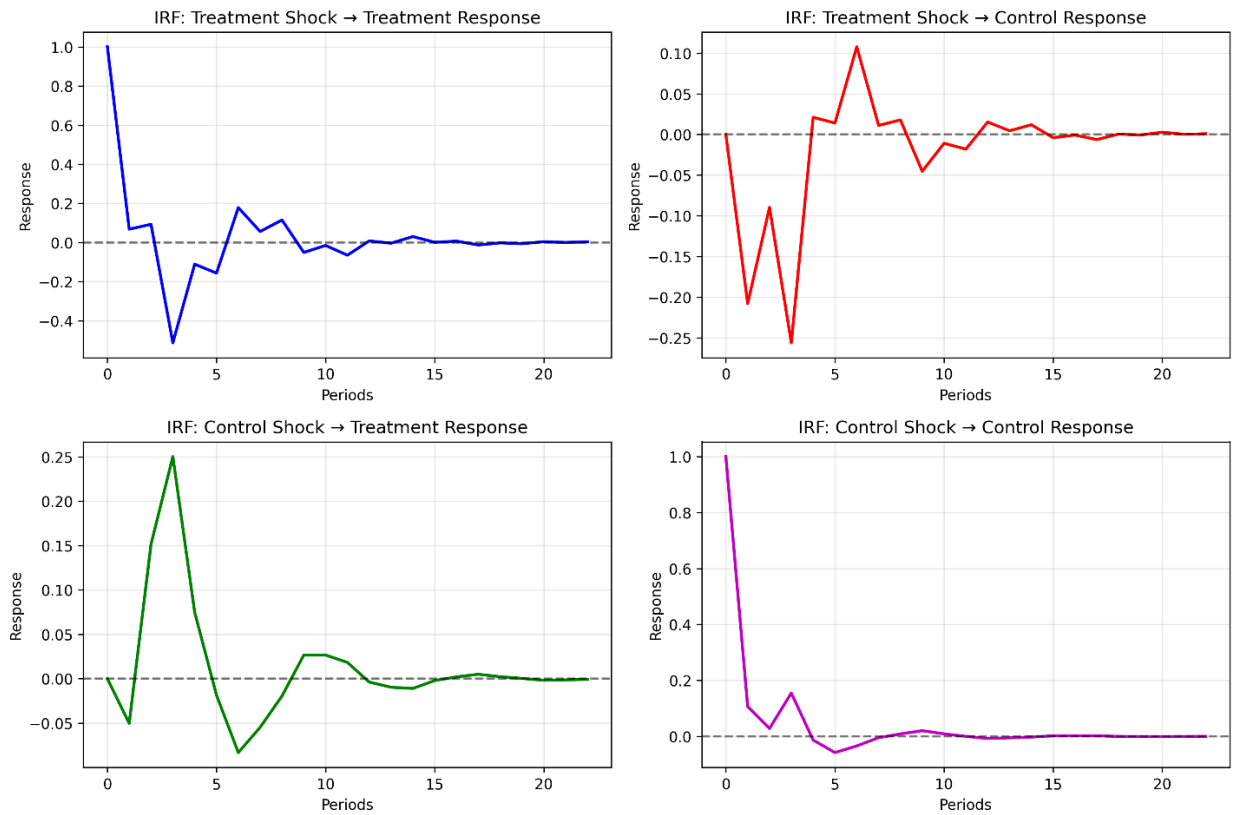

**Figure 5:** IRF results of the MUFJ M&A event on October 3, 2005

Note: The panels illustrate the dynamic response of each portfolio to a one-standard-deviation shock in the other. The key finding is in the top-right panel, which shows the response of the control portfolio to a shock in the treatment portfolio. The plot indicates a volatile but ultimately positive spillover; a positive shock to the merging banks' returns is met with an initial negative market reaction in competitor returns, followed by a strong positive response peaking around the fifth period, before closing at zero. The bottom-left panel shows a significant response of the treatment portfolio to a shock in the control portfolio, with a peak around the third period. The remaining panels show each portfolio's expected, rapidly decaying response to a shock.

**Table 3:** Multi-Period Granger Causality Test Results for Resona Holdings M&A

| Lags Tested | Treatment → Control | | Control → Treatment | |
|---|---|---|---|---|
| | F-statistic | p-value | F-statistic | p-value |
| 1 | 0.18 | 0.67 | 0.31 | 0.58 |
| 2 | 3.8 | 0.02** | 1.85 | 0.16 |
| 3 | 2.31 | 0.07* | 0.81 | 0.49 |
| 4 | 1.14 | 0.34 | 0.32 | 0.86 |
| 5 | 1.08 | 0.37 | 0.47 | 0.79 |
| 6 | 0.97 | 0.45 | 0.52 | 0.80 |

Note: The results indicate a unidirectional spillover from the treatment to the control portfolio. The chosen lag length for the final model was 2. Tests at other lag lengths (1, 3-6) did not show significant spillovers in either direction, except a marginally significant effect (p=0.078) from Treatment → Control at lag 3.

**Table 4:** Multi-Period Granger Causality Test Results for the MUFG M&A

| Lags Tested | Treatment → Control | | Control → Treatment | |
|---|---|---|---|---|
| | F-statistic | p-value | F-statistic | p-value |
| 1 | 4.67 | 0.03** | 0.05 | 0.82 |
| 2 | 2.56 | 0.08* | 2.33 | 0.10 |
| 3 | 3.26 | 0.02** | 1.75 | 0.16 |
| 4 | 2.38 | 0.05* | 1.35 | 0.26 |
| 5 | 1.78 | 0.12 | 1.66 | 0.15 |
| 6 | 1.38 | 0.23 | 2.15 | 0.05* |
| 7 | 1.09 | 0.38 | 2.05 | 0.05* |

Note: The table shows F-statistics and corresponding p-values for Granger causality tests performed at different lag lengths. The test for Treatment → Control causality is statistically significant at multiple lags.

## 3.3 PSM-ATT Results

The PSM in Table 5 matched the 43 daily observations of the portfolio of the Resona Holdings M&A in the pre-event period to a control group (non-M&A banks) based on observable characteristics like lagged returns, moving averages, and volatility. The matching process was highly successful, finding comparable counterparts for 37 of the 43 observations (an 86% match rate) in the M&A portfolio, which creates a strong basis for comparison. Table 5 reports a positive and highly statistically significant ATT. This result can be understood by comparing the post-event performance of the groups: the average daily return for the treatment portfolio was 0.20% while the matched control group had an average daily return of just 0.02%. This difference represents the estimated causal impact of the merger, providing stronger, causally-informed evidence that complements the CAR findings and suggests a sustained positive effect on the firms' performance.

The MUFG M&A appears to have been a unique event, making it challenging to find comparable counterparts among its peers. The propensity scores for the treatment group were heavily skewed towards 1 (mean of 0.69) in Table 6, while the control group's scores were skewed towards 0 (mean of 0.31). This finding is consistent with the extraordinary CAR results, which showed a massive +29% abnormal return. It seems the PSM algorithm struggled to find control banks with a similarly high probability of "treatment" because no other banks in the sample exhibited the unique combination of size, performance, and strategic positioning that led to the MUFG deal.

**Table 5**. PSM and ATT Results for the Resona M&A

| Metric | Value | Component Means | Value |
| --- | --- | --- | --- |
| ATT Estimate | 0.00186 | Mean Return (Treatment Group) | 0.00204 |
| Bootstrap Standard Error | 0.00035 | Mean Return (Matched Control Group) | 0.00019 |
| T-statistic | 5.28*** | **Matching Quality** | |
| Bootstrap 95% Confidence Interval | [0.0012, 0.0026] | Number of Matched Pairs | 37 |

Note: The ATT is estimated on a matched sample created using caliper matching on propensity scores. Covariates for the propensity score model included lagged returns, a 3-day moving average

of returns, and return volatility. The ATT represents the average daily return difference between the treatment portfolio and its matched counterfactual.

Table 6. PSM and ATT Results for the MUFG M&A

| Metric | Value | Component Means | Value |
|---|---|---|---|
| ATT Estimate | -0.00086 | Mean Return (Treatment Group) | 0.00178 |
| Bootstrap Standard Error | 0.00101 | Mean Return (Matched Control Group) | 0.00265 |
| T-statistic | -0.82 | **Matching Quality** | |
| Bootstrap 95% Confidence Interval | [-0.0028, 0.0012] | Number of Matched Pairs | 20 |

Note: The ATT is estimated using a matched sample created using caliper matching. The low match rate (47.6%) suggests a lack of common support between the treatment and control groups. Therefore, the ATT estimate should be interpreted cautiously as it is based on a non-random subset of the treatment observations.

### 3.4. Robustness Tests

Several additional tests were performed to check the reliability of the main findings. First, this study performed a placebo test using the complete event study method on a randomly chosen date with no merger news. The Granger causality results were then examined for sensitivity to the model's lag length. Finally, the PSM quality was confirmed by checking for covariate balance. The results of robustness tests on two M&A events are provided in the Appendix.

### 3.5. Discussion and Implications

The event studies for the 2005 MUFG formation and the 2018 Resona merger reveal strongly positive and statistically significant market reactions, confirming that investors viewed both M&A as value-creating. However, the economic magnitude of the market reaction to the MUFG M&A was much larger, with a CAR of approximately +29% compared to about +6% for the Resona M&A.

The Granger causality tests for the Resona and MUFG mergers reveal a statistically significant and unidirectional information spillover from the merging banks to their competitors, with no evidence of a reverse effect. However, the nature of this spillover appears to differ: the Resona merger was associated with a clean, short-lived spillover at a two-day lag. In contrast, the larger MUFG merger showed a more complex and persistent predictive relationship at multiple lags. This suggests that while both events transmitted information to the broader market, the more transformative MUFG deal triggered a more prolonged period of market reassessment among competitor firms.

The PSM analysis produced strikingly different outcomes for the two mergers. The method proved effective for the 2018 Resona M&A, indicating a sustained and positive causal impact. In contrast, the analysis of the 2005 MUFG M&A showed a key limitation of the method. A low match rate indicated that the merging banks were unique, making it difficult to find a credible control group. This divergence shows that while PSM can be a powerful tool, its effectiveness appears to depend on the comparability of the event to others in the market.

This multi-method approach has its limitations. PSM is still affected by the unobserved factors and outliers. Likewise, Granger causality identifies statistical predictability, not necessarily economic causation. However, finding consistent results across these different methods would increase confidence in the conclusions.

## 4. Conclusion

This study examined the market's reaction to two major bank mergers in Japan: one following the Financial Big Bang and the other after the global financial crisis. By analyzing the merger effects on the banks involved and comparable banks, the study demonstrates that investors viewed both events as significant value-creating opportunities. The analysis combined event study techniques with time-series and matching methods to deconstruct the impact into an immediate wealth effect, a dynamic spillover process, and a more refined causal estimate. The results show that while both mergers were received favorably, the market reaction to the globally transformative MUFG deal was substantially larger than the more regionally focused Resona consolidation.

Furthermore, the findings indicate that these merger events created apparent, unidirectional information spillovers to competitor firms. The impact of these spillovers appeared to depend on the scale of the event; the Resona M&A triggered a quick, transient market adjustment, while the MUFG M&A led to a more complex and prolonged reassessment of the competitive landscape. Unlike most other M&A events, which suppress competitors' performance or result in the superior performance of competitors due to distrust of the merging firms, these mergers appeared to generate synergistic effects.

## Acknowledgment

This work was supported by the University Research and Development Awards program at Texas A&M International University.

**Appendix**

**A. Event Study Models to Estimate CAR**

CAR measures the immediate wealth effect. An abnormal return $(AR_{i,t})$ on day $t$ is the difference between the actual return of a firm $i$ $(R_{i,t})$ and its expected "normal" return $(E[R_{i,t}])$.

$$AR_{i,t} = R_{i,t} - E[R_{i,t}] \tag{A.1}$$

The normal return is estimated using the market model, which regresses the stock's return on the market return over the estimation window. The CAR is the sum of the daily ARs over the M&A event window $[T_1, T_2]$:

$$CAR_i(T_1, T_2) = \sum_{t=T_1}^{T_2} AR_{i,t} \tag{A.2}$$

Statistical significance is assessed with a t-test with the standard deviation of the residuals during the estimation period. This study implements three models for the event study.

The *market model* provides a straightforward statistical analysis of how an individual asset's return relates to the market's returns. It is an empirical tool that captures systematic market risk by separating a security's return into two pieces. One part is the systematic return, driven by broad market trends. The other is the idiosyncratic return, which comes from factors specific to the bank in the M&A event. The model uses a simple linear regression:

$$R_{i,t} = \alpha_i + \beta_i R_{m,t} + \varepsilon_{i,t} \tag{A.3}$$

where the expected return = $\hat{\alpha}_i + \hat{\beta}_i R_{m,t}$. $R_{i,t}$ is the return of asset $i$ in period $t$; $R_{m,t}$ is the return of the overall market portfolio in period $t$; $\alpha_i$ is the regression intercept, representing the asset's average return that isn't explained by the market's movement; $\beta_i$ represents the asset's sensitivity to market returns, serving as a measure of its systematic risk; $\epsilon_{i,t}$ is the error term, capturing the idiosyncratic return for asset $i$ in period $t$.

The *CAPM* assumes that an asset's expected return should be the risk-free rate plus a premium for taking on systematic risk in an efficient market. For event studies, the CAPM is often expressed as a regression of excess returns (returns above the risk-free rate):

$$R_{i,t} - R_{f,t} = \alpha_i + \beta_i(R_{m,t} - R_{f,t}) + \varepsilon_{i,t} \tag{A.4}$$

where, $R_{i,t} - R_{f,t}$ is the excess return of asset $i$ at time $t$; $(R_{m,t} - R_{f,t})$ is the market risk premium factor; $\beta_i$ represents the asset's sensitivity to market returns, serving as a measure of its systematic risk; $\epsilon_{i,t}$ is the error term, capturing the idiosyncratic return for asset $i$ in period $t$. $\alpha_i$ should be zero for any correctly priced asset.

The *Fama-French three-factor model* suggested that two other factors in CAMP seemed to be consistently priced by the market. In addition to market risk, the *Fama-French three-factor* model considers the size and value premiums. The model, therefore, provides a more demanding benchmark for performance evaluation and for calculating an asset's "normal" return. It is specified as a multiple regression:

$$R_{i,t} - R_{f,t} = \alpha_i + \beta_{i,mkt}(R_{m,t} - R_{f,t}) + \beta_{i,smb}\text{SMB}_t + \beta_{i,hml}\text{HML}_t + \varepsilon_{i,t} \quad (A.5)$$

where $R_{i,t} - R_{f,t}$ is the excess return of asset $i$ at time $t$; $(R_{m,t} - R_{f,t})$ is the market risk premium factor; $SMB_t$ is the size premium, representing the excess return of small-cap stocks over large-cap stocks at time $t$; $HML_t$ is the value premium, representing the excess return of value stocks over growth stocks at time $t$; $\beta_{i,mkt}, \beta_{i,smb}, \beta_{i,hml}$ are the factor sensitivities, or loadings, for asset $i$. For example, a positive $\beta_{i,smb}$ indicates the stock tends to behave more like a small-cap stock. $\epsilon_{i,t}$ is the random error term, representing the return unexplained by the three factors. $\alpha_i$ represents the return portion that is not explained by any of the three factors, and it provides strong evidence of abnormal performance when its value is statistically different from zero.

## B. Granger Causality and IRF: Detecting Spillovers

To test for information spillovers between the treatment and control portfolios, a bivariate VAR model of order $p$ is estimated:

$$R_t^{treat} = \alpha_1 + \sum_{i=1}^{p}\beta_{1,i}R_{t-i}^{treat} + \sum_{i=1}^{p}\gamma_{1,i}R_{t-i}^{control} + \epsilon_{1,t} \quad (B.1)$$

$$R_t^{control} = \alpha_2 + \sum_{i=1}^{p}\beta_{2,i}R_{t-i}^{treat} + \sum_{i=1}^{p}\gamma_{2,i}R_{t-i}^{control} + \epsilon_{2,t} \quad (B.2)$$

A Granger causality test is used in this analysis to examine the joint significance of the lagged coefficients. Rejecting this null hypothesis implies that the control group's returns help predict the treatment group's returns. Impulse response functions (IRFs) from the VAR are also analyzed to see how a shock to one portfolio's returns affects the other over time. The lag length $p$ is chosen using information criteria like AIC or BIC.

## C. PSM and ATT: Estimating Causal Effects

A main challenge in evaluating M&A is that the M&A decision, driven by regulatory pressures and strategic goals, is a non-random selection. The prolonged low-profitability environment in Japan drove banks to seek merger partners (Montgomery, 2005). Such a non-random selection can bias the results of simpler comparative methods (Baker et al., 2025; Goodman-Bacon, 2021). The goal is to build a control group of non-merging banks that, based on observable characteristics, closely resembles the group of merging banks before the event.

The PSM procedure begins by modeling a bank's probability (propensity score) in the "treatment group" using a logistic regression. The covariates are based on performance metrics like past stock returns, risk metrics like stock return volatility (Hagendorff & Keasey, 2009), and size and business model measures, which could influence the merger decision and future returns. Following this, each treated bank is matched to a control bank with a similar propensity score. Instead of trying to match on many different variables simultaneously, PSM calculates the propensity score. For any given bank, the propensity score is its estimated probability of being part of the treatment group (M&A event).

$$p(X_i) \equiv P(D_i = 1|X_i) \tag{C.1}$$

Here, $D_i$ represents whether bank $i$ received the treatment, and $X_i$ is its vector of pre-treatment characteristics. In practice, this score is usually estimated with a logistic or probit model where the treatment status is the outcome variable. By creating a comparable control group, PSM provides a way to assess this unobservable counterfactual. The ATT can then be calculated from the sample data by taking the average difference in outcomes between the treated firms and their matched control counterparts.

$$\widehat{ATT} = \frac{1}{N_T} \sum_{i \in \{D=1\}} \left( Y_i - \hat{E}[Y_j(0) | i \text{ is matched to } j] \right) \tag{C.2}$$

In this calculation, $N_T$ is the number of banks in the M&A event, $Y_i$ is the observed outcome for bank $i$ in the M&A event, and $\hat{E}[Y_j(0)]$ is the average outcome of the non-M&A bank $j$ that was matched to bank $i$.

## D. Robustness Test

**Figure D1:** Cross-Correlation between Treatment and Control Portfolio Returns for Resona Holdings M&A on April 1, 2018

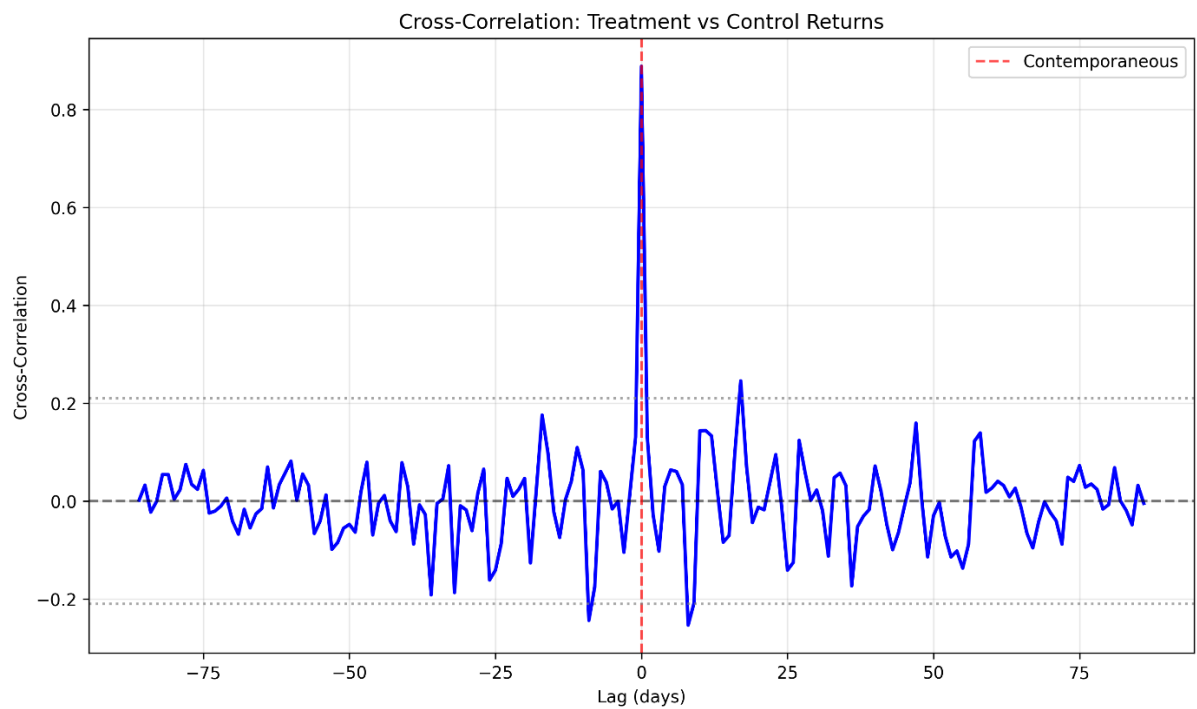

Note: This figure displays the cross-correlation function between the daily returns of the treatment and control portfolios for up to 90 days of leads and lags. The function reveals a single, overwhelmingly dominant spike at a lag of zero (contemporaneous). The peak cross-correlation at this point is 0.89, indicating a very strong positive relationship between the two portfolios on the same trading day. This strong contemporaneous correlation is expected, as both portfolios are composed of Japanese banking stocks that are subject to the same daily market-wide and sector-wide news and sentiment.

The absence of significant spikes at other lags is consistent with the findings from the Granger causality and impulse response function analyses. While the Granger test identified a statistically significant predictive relationship from the treatment to control group at a two-day lag, its economic magnitude was small and transient. The cross-correlation plot reinforces this by showing that, in comparison to the powerful same-day relationship, any lead-lag effects are minor. This pattern suggests that while a subtle, short-lived spillover effect exists, the primary dynamic linking the two portfolios is their shared, immediate reaction to common information.

**Figure D2:** Cross-Correlation between Treatment and Control Portfolio Returns for the MUFJ M&A event on October 3, 2005

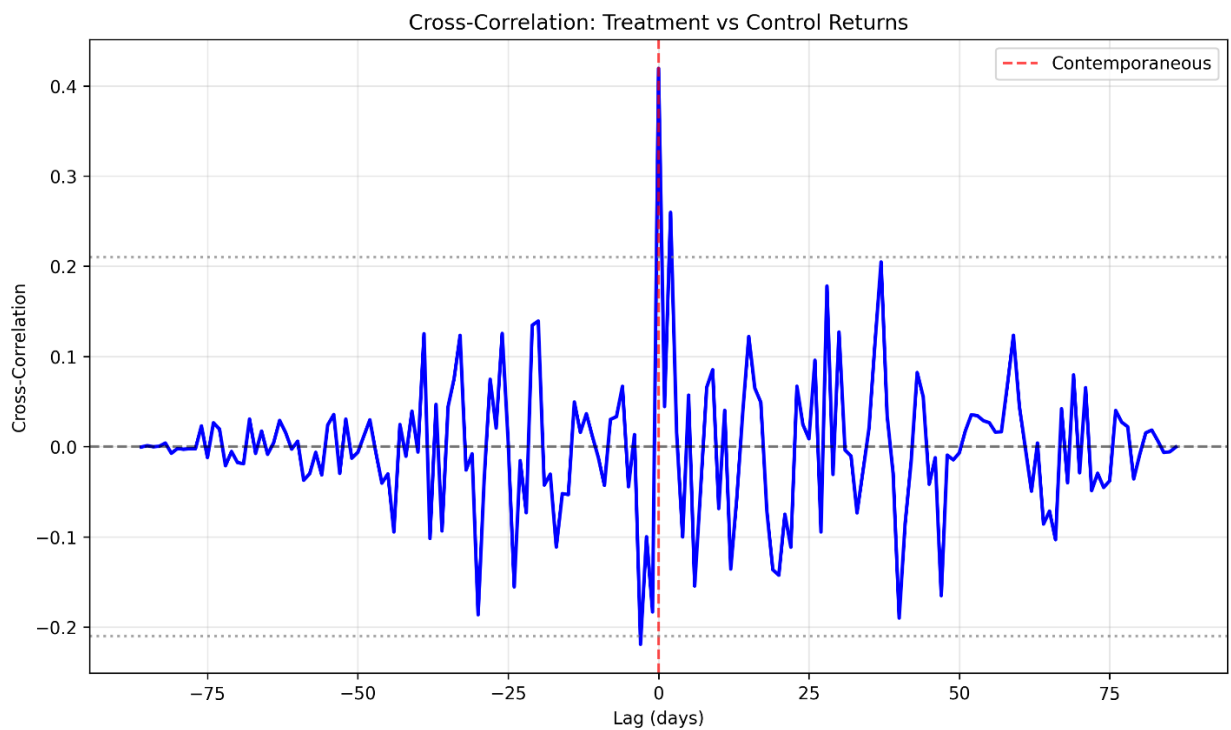

Note: This figure displays the cross-correlation function between the daily returns of the treatment (MUFG) and control portfolios. The plot shows a distinct and statistically significant positive spike at a lag of zero (contemporaneous), with a peak correlation of 0.42. This indicates a moderately strong tendency for the merging banks and their competitors to move in the same direction on the same trading day, which is expected given their shared exposure to common market and industry-wide factors.

The plot also reveals several smaller, though still significant, positive correlations at various lags and leads, which is consistent with the more complex dynamic relationship identified in the Granger causality and IRF analyses. Unlike a simple, clean spillover, the presence of multiple significant cross-correlations suggests a more persistent and interactive relationship between the merging firms and their peers during this transformative period. However, the dominant feature remains the strong contemporaneous link, indicating that the primary driver of the relationship is the shared reaction to daily information.

**Results of Robustness Tests**

**Table D1: Summary of Robustness Tests for the Resona M&A Analysis**

| Test | Method | Outcome Metric | Result | Interpretation |
| --- | --- | --- | --- | --- |

| Test | Method | Outcome Metric | Result | Interpretation |
| --- | --- | --- | --- | --- |
| Placebo Event Study | CAR | p-value | -0.09% (0.937) | **Pass:** No effect on a random date. |
| Granger Causality | VAR Lag Sensitivity | p-value (T → C) | Lag 2: 0.025**<br>Lag 3: 0.078* | **Pass:** Spillover is stable to lag choice. |
| PSM | Covariate Balance | p-value | 0.950 | **Pass:** Matched groups are comparable. |

Note: The placebo test applies the event study methodology to a randomly selected non-event date. The results showed no significant abnormal returns, with a final CAR of only -0.46% (t-statistic = -0.98). This helps confirm that the strong positive effect measured on the true event date was likely not a random occurrence. The Granger causality test checks if the spillover finding holds for different VAR model lag lengths. While the main analysis used a lag of 2, the spillover from the treatment to the control group remained statistically significant at the 5% level for a lag of 2 and at the 10% level for a lag of 3, which suggests the finding is reasonably stable. The covariate balance test confirms that the PSM procedure successfully created a comparable control group by ensuring no significant differences in pre-treatment characteristics. After matching, the standardized mean differences for all pre-treatment measures were well below accepted thresholds, and t-tests showed no significant differences between the treated and matched control groups. This successful balancing provides credibility to the causal interpretation of the estimated ATT.

**Table D2: Summary of Robustness Tests for the MUFG M&A Analysis**

| Test | Method | Outcome Metric | Result | Interpretation |
| --- | --- | --- | --- | --- |
| Placebo Event Study | CAR | p-value | -1.14% (0.356) | **Pass:** No effect on a random date. |
| Granger Causality | VAR Lag Sensitivity | p-value (T → C) | Lag 1: 0.032**<br>Lag 3: 0 | **Pass:** Spillover is stable to lag choice. |
| PSM | Covariate Balance | p-value | 0.784 | **Pass:** Matched groups are comparable* |

Note: The placebo test applies the event study methodology to a randomly selected non-event date. The results showed no significant abnormal returns, with a final CAR of only -0.46% (t-statistic = -0.98). This helps confirm that the strong positive effect measured on the true event date was likely not a random occurrence. The Granger causality test checks if the spillover finding holds for different VAR model lag lengths. While the main analysis used a lag of 2, the spillover from the treatment to the control group remained statistically significant at the 5% level for a lag of 2 and

at the 10% level for a lag of 3, which suggests the finding is reasonably stable. *These PSM checks apply only to the 47.6% of treatment observations for which a comparable match could be found and should be interpreted with caution.